\documentclass[twocolumn,showpacs,preprintnumbers,amsmath,amssymb,aps]{revtex4}

\usepackage{graphicx}% Include figure files
\usepackage{hyperref}
\begin{document}

%\preprint{Submitted to PRL, Oct. 26, 2004}

\title{Two-photon coincident-frequency-entanglement via extended phase matching}

\author{Onur Kuzucu, Marco Fiorentino, Marius A. Albota, Franco N. C. Wong,
and Franz X. K\"{a}rtner}
%\email{xxx@mit.edu}
\affiliation{Research Laboratory of Electronics, Massachusetts
Institute of Technology, Cambridge, MA 02139}

\begin{abstract}
We demonstrate a new class of frequency-entangled states generated
via spontaneous parametric down-conversion under extended phase
matching conditions. Biphoton entanglement with coincident signal
and idler frequencies is observed over a broad bandwidth in
periodically poled KTiOPO$_4$.  We demonstrate high visibility in
Hong-Ou-Mandel interferometric measurements under pulsed pumping
without spectral filtering, which indicates excellent frequency
indistinguishability between the down-converted photons. The
coincident-frequency entanglement source is useful for quantum
information processing and quantum measurement applications.
\end{abstract}


\maketitle

Improvement in measurement accuracies beyond the classical limits
is one of the most attractive applications of quantum
entanglement. For example, one can perform two-photon
interferometric measurements with sensitivity better than the
diffraction limit by exploiting the momentum entanglement of
photon pairs generated from spontaneous parametric down-conversion
(SPDC) \cite{others}. Recently, Giovannetti {\em et al.}
\cite{lloyd-nature} suggested that timing and positioning
measurements can be enhanced by $\sqrt{N}$ over the standard
quantum limit $\Delta t = 1/\sqrt{N}\Delta\omega$  if one uses the
$N$-photon coincident-frequency-entangled state, with mean
frequency $\overline\omega$, $| \psi \rangle = \int d\omega
\phi(\omega) | \overline\omega + \omega \rangle_1...|
\overline\omega + \omega \rangle_N$, where $\phi(\omega)$ is the
spectral amplitude, $\Delta t$ is the timing accuracy, and
$\Delta\omega$ is the coherence bandwidth of each photon.  For the
simplest nontrivial case of $N=2$ the
coincident-frequency-entangled state $| \psi \rangle_{cf} = \int
d\omega \phi(\omega) | \overline\omega + \omega \rangle_1 |
\overline\omega + \omega \rangle_2$ consists of a pair of
entangled photons with identical but uncertain frequencies; the
two photons are positively correlated in frequency, and hence
anti-correlated in time.  In contrast, conventional SPDC, the most
common method for generating entangled photon pairs, typically
yields pairs of time-entangled photons that are correlated in time
but anti-correlated in frequencies as described by the state $|
\psi \rangle_{spdc} = \int d\omega \phi(\omega) | \overline\omega
+ \omega \rangle_1 | \overline\omega - \omega \rangle_2$.  As a
result, SPDC under conventional phase matching condition is not
suitable for generating coincident-frequency entanglement.

The ability to generate a coincident-frequency entangled state also
provides a solution to the problem of frequency distinguishability
in pulsed SPDC. This problem is particularly relevant when pulsed
pumped SPDC is used to increase temporal resolution in timing or
other measurement applications. Because of the large pump bandwidth
and frequency anti-correlation in the down-converted light, pulsed
SPDC typically generates photon pairs with signal and idler
frequency spectra that are not identical \cite{theory}. This
spectral distinguishability under pulsed pumping causes
Hong-Ou-Mandel (HOM) interferometric measurements \cite{HOM} to have
poor visibility \cite{expt}. Coincident-frequency entanglement, on
the other hand, yields photon pairs with the same spectral
properties, thereby restoring the HOM visibility.   This type of
entanglement can be used in a broad variety of applications ranging
from linear optics quantum computing (LOQC) \cite{KLM} to timing
measurements \cite{lloyd-nature}.

We note that pulsed polarization entanglement with high visibility
has been previously demonstrated in a dual-port SPDC configuration
that separates the signal and idler fields at the output
\cite{Shih}. Because the signal and idler fields are not mixed,
their polarization entanglement is not sensitive to the difference
in their spectra, a concept that is used in bidirectionally pumped
SPDC with interferometric combination \cite{dual-pump}.  However,
such techniques would still show low HOM visibility under pulsed
pumping because the spectra of the interfering photons are
different.

Several groups recently suggested the possibility of generating a
biphoton output that approximates the two-photon
coincident-frequency entangled state $|\psi\rangle_{cf}$ by
engineering the SPDC phase-matching function
\cite{walmsley,sasha,epmPRL,epmPRA}. In this Letter we report on
the experimental generation of two-photon coincident-frequency
entanglement by use of extended phase matching (EPM) in type-II
phase-matched periodically poled KTiOPO$_4$ (PPKTP), as proposed
by Giovannetti {\em et al.} \cite{epmPRL,epmPRA}.  We obtain high
HOM visibility under pulsed pumping with a pulse width shorter
than the biphoton coherence time. The EPM technique eliminates the
degradation in HOM visibility associated with a long crystal under
pulsed pumping. This technique should lead to better entanglement
sources for quantum applications such as cryptography,
teleportation, and quantum computing, as well as timing
measurements.

For entanglement generation via SPDC, one is mostly interested in
the frequency degenerate case in which the signal and idler
frequencies are equal to half of the pump frequency.  Therefore,
we assume degenerate collinear down-conversion at pump frequency
$\omega_p$ with the phase mismatch $\Delta k \equiv k_p(\omega_p)
- k_s(\omega_p/2) - k_i(\omega_p/2)= 0$, where $k_{p,s,i}$ are the
moduli of the wavevectors in the crystal for the pump ($p$),
signal ($s$), and idler ($i$). When the pump frequency is tuned,
$\Delta k$ is no longer zero for a frequency degenerate output.
This can be rectified if we operate at a point where the phase
mismatch's first frequency derivative $\Delta k' =
\partial (\Delta k)/\partial \omega$ is also zero, which yields
$k'_p(\omega_p) = (k'_s(\omega_p/2) + k'_i(\omega_p/2))/2$. The
conditions $\Delta k =0$ and $\Delta k' = 0$ are called extended
phase matching \cite{epmPRL,epmPRA}. The consequence of EPM is
that the phase-matched signal and idler frequencies remain equal
as $\omega_p$ is tuned or if a pulsed pump is used.  This remains
true if the pump tuning range or the pump bandwidth $\Omega_p$ is
smaller than the extended phase matching bandwidth $\Omega_{epm}$,
which is determined by the second frequency derivative of $\Delta
k$ \cite{epmPRA}. Because $k'$ is the inverse of the group
velocity EPM is also known as zero group velocity mismatch in
ultrafast nonlinear optics.

Our coincident-frequency entanglement generation method is
conceptually simple.  In addition to operating under EPM
conditions, we use a long crystal of length $L$ to narrow the
phase matching bandwidth whose full width at half maximum is
\begin{equation}
\Omega_c = \frac{2\pi}{|k'_s - k'_i|L}\,.
\end{equation}
Type-II phase matching is used so that $k'_s \neq k'_i$ and a
small $\Omega_c$ can be obtained for coincident-frequency output
$\omega_s \approx \omega_i$. The phase matching bandwidth
$\Omega_c$ is related to the biphoton coherence time $\tau_c =
2\pi/\Omega_c = |k'_s - k'_i|L$. The coincident-frequency
entangled state $|\psi\rangle_{cf}$ is generated if \cite{epmPRA}
\begin{equation}
\Omega_c \ll \Omega_p \ll \Omega_{epm}\,,
\end{equation}
which ensures that the HOM visibility is unaffected by pulsed
pumping.

We have chosen a 10-mm long PPKTP crystal for coincident-frequency
generation.  Periodic poling in PPKTP allows one to use the
grating period as a free parameter for phase matching at any set
of operating wavelengths within the crystal's transparency window.
Thus, we can always choose a grating period to satisfy the
effective conventional phase matching condition $\Delta k = 0$.
For noncritical type-II  phase matching with collinear propagation
along the crystal's $x$-axis, $\Delta k'$ is entirely determined
by the dispersion of the crystal.  Recently we have experimentally
determined that for a pump wavelength $\lambda_p \approx 792$ nm
the EPM conditions for degenerate collinear type-II phase matching
are satisfied with a grating period of $\sim$46.15 $\mu$m
\cite{epmSHG}.  Using this PPKTP crystal we demonstrated efficient
ultrabroadband SHG over an EPM bandwidth of 67\,nm centered at the
fundamental wavelength of $\sim$1584\,nm.  This bandwidth is
nearly two orders of magnitude larger than typical SHG bandwidths.

\begin{figure}[tb]
\centerline{\includegraphics[width=3.2in]{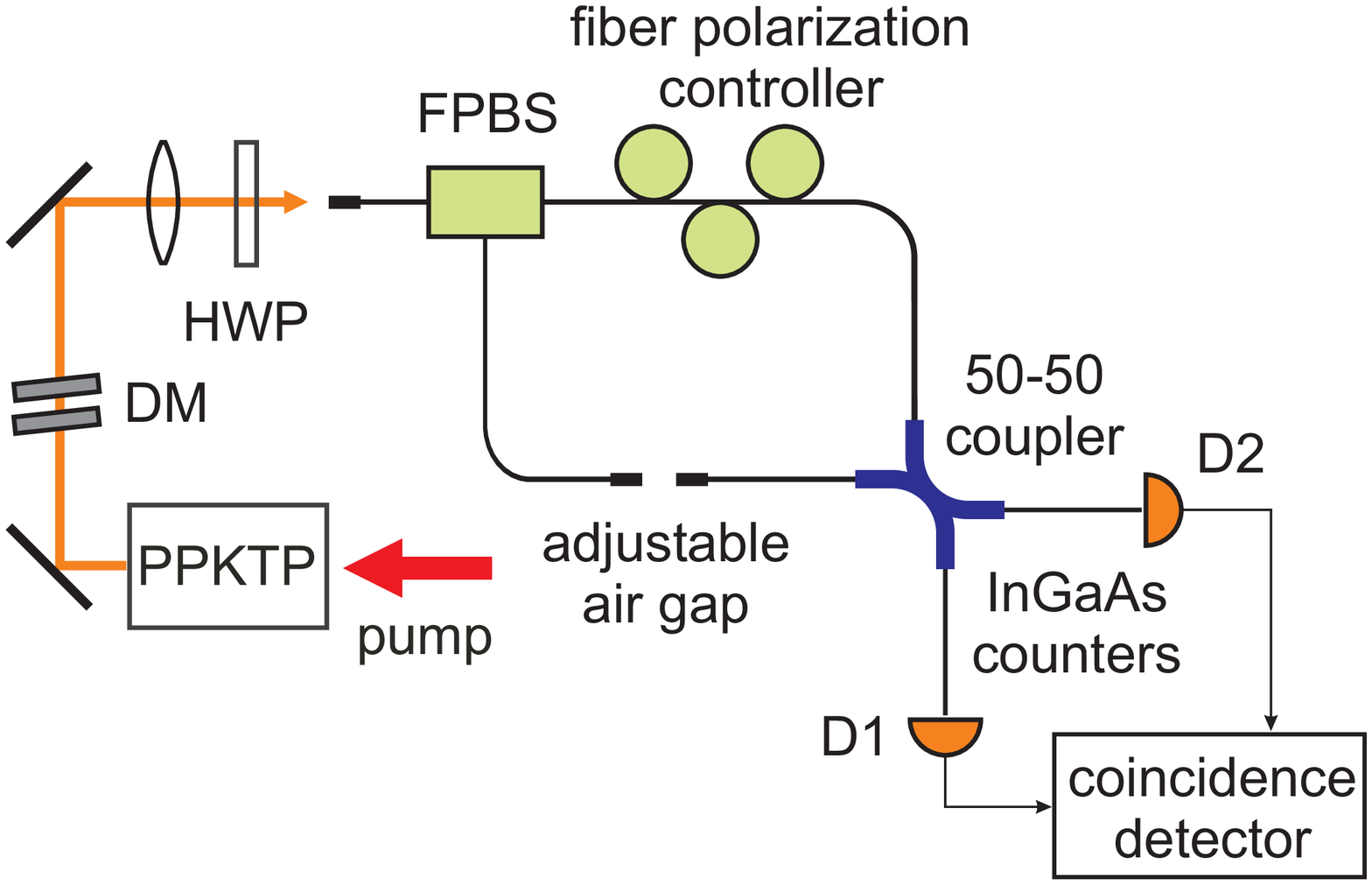}}
\caption{Experimental setup showing the entanglement source and
the fiber HOM interferometer used to test the
coincidence-frequency entangled state. DM: dichroic mirror; HWP:
half-wave plate; FPBS: fiber polarization beam splitter.}
\label{setup}
\end{figure}

Figure \ref{setup} shows the experimental setup for the generation
of coincident-frequency entanglement and the HOM interferometric
analysis. The pump was a Ti:sapphire laser (Spectra-Physics,
Tsunami) that could be operated in either continuous-wave (cw) or
pulsed mode without altering its output spatial mode
characteristics.  In pulsed operation, the mode-locked laser
output had a 3-dB bandwidth of $\sim$6\,nm and an average power of
350 mW at an 80-MHz pulse repetition rate. The PPKTP crystal was
antireflection coated at 792 and 1584 nm and the pump light was
focused into it with a diameter of $\sim$200\,$\mu$m.  The output
was collimated and two dichroic mirrors (reflecting the 792\,nm
pump and transmitting the 1584\,nm signal and idler) were used to
eliminate the pump and no narrowband spectral filtering was used
to restrict the output bandwidth. The resultant output was then
focused into a single-mode polarization-maintaining (PM) optical
fiber for subsequent HOM interferometric measurements.

The fiber-coupled light was sent to a fiber polarizing beam
splitter (FPBS) that separated the orthogonally polarized signal
and idler photons into their respective fiber channels.  A
half-wave plate (HWP) at the entrance to the PM fiber was used to
set up the signal and idler in their appropriate fiber
polarization modes. In the signal arm of the HOM interferometer
was a fiber polarization rotator that was used to match the idler
polarization in the second arm. The two arm lengths between the
FPBS and the 50--50 fiber coupler were carefully matched by
including an adjustable air gap in the idler arm.  The air gap was
nominally 50\,mm-long and efficient coupling ($>$70\%) between the
two fibers was achieved by attaching a collimator to each fiber.

The two outputs of the 50--50 coupler were sent to two
fiber-coupled custom-made InGaAs single-photon counters. The
design and performance of similar devices have been described in
more detail in our previous work \cite{InGaAs,InGaAs2}. The InGaAs
detectors were operated in Geiger mode by simultaneously sending a
20-ns gating pulse (3.9\,V above the breakdown voltage) to each
detector at a repetition rate of 50 kHz, yielding a detector duty
cycle of 10$^{-3}$. The counter D1 (D2) had a detection quantum
efficiency of 16\% (24\%) at $\sim$1580\,nm. Under the above
operating conditions we measured dark count rates of 40/s and 20/s
for D1 and D2, respectively, which correspond to dark count
probabilities of $8\times 10^{-4}$ (D1) and $4\times 10^{-4}$ (D2)
per gate. The photocount outputs were amplified and sent into a
high-speed AND-gate logic circuit for coincidence detection within
a 1.8-ns coincidence window.  For cw pumping with 350 mW pump
power, we have observed average singles count probabilities of
$2.3\times 10^{-3}$ (D1) and $2.5\times 10^{-3}$ (D2) per 20-ns
gate, corresponding to count rates of 125/s and 115/s respectively
(not corrected for dark counts). The probability of detecting a
pair per gate is $2\times 10^{-5}$ corresponding to a count rate
of 1 coincidence/s. The low coincidence count rate is due to the
interferometer losses ($\sim$65\% transmissivity, measured
independently), low detector quantum efficiencies, and an
estimated 30\% coupling efficiency into the PM fiber. From the
detection efficiencies and our measurement duty cycle we estimate
a single spatial fiber-optic mode pair production rate of $\sim$4
$\times 10^6$/s at 350\,mW of pump power.

\begin{figure}[tb]
\centerline{\includegraphics[width=3.2in]{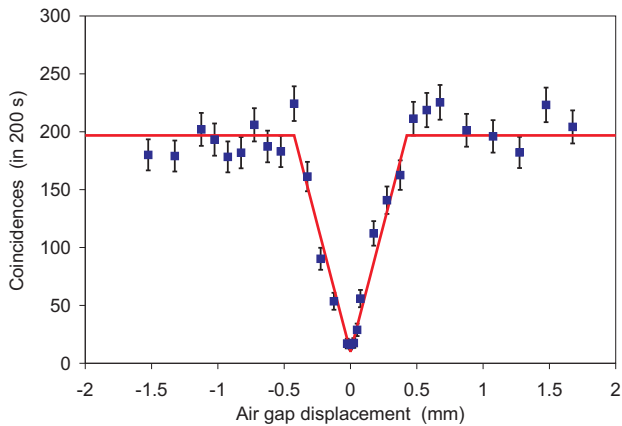}} \caption{HOM
interferometric measurements under cw pumping at 792\,nm. The
experimental data are fitted with a triangular-dip HOM function
(solid line). Base-to-base width of the HOM dip is $l_c = 0.85 \pm
0.16$\,mm corresponding to a biphoton coherence time $\tau_c =
1.42 \pm 0.26$\,ps. HOM dip visibility is $95 \pm 5$\%.}
\label{cwhom}
\end{figure}

\begin{figure}[tb]
\centerline{\includegraphics[width=3.2in]{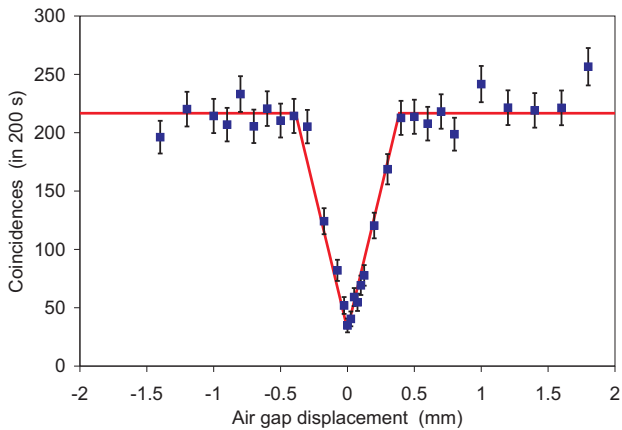}}
\caption{HOM interferometric measurements for a pulsed pump
centered at 790 nm  with a 3-dB bandwidth of 6 nm. The
experimental data are fitted with a triangular-dip HOM function
(solid line). Base-to-base width of the HOM dip is $l_c = 0.8 \pm
0.2$\,mm corresponding to a biphoton coherence time $\tau_c = 1.3
\pm 0.3$\,ps. HOM dip visibility is $85 \pm 7$\%.}
\label{pulsedhom}
\end{figure}

For the HOM measurements, the coincidence counts and the two
singles counts were measured as a function of the air gap
distance.  Accidental coincidence counts were separately measured
by using the same technique described in Refs.\
\cite{InGaAs,InGaAs2}. Typical accidental coincidence
probabilities within the 1.8-ns coincidence window were $5\times
10^{-7}$ per gate corresponding to 5 counts in a 200-s measurement
time.  Figures \ref{cwhom} and \ref{pulsedhom} show the data
corrected for the accidental counts.

Figure~\ref{cwhom} shows the HOM measurements under cw pumping at
the pump wavelength of 792\,nm, and we obtained similar results at
a different pump wavelength of 787\,nm. The biphoton coherence
time is related to the base-to-base width of the HOM dip $l_c$ by
$\tau_c = l_c/2 c$. Fitting the experimental data with the HOM
triangular dip expected for type-II phase matching yields a
visibility of $95 \pm 5$\% and a biphoton coherence time of $1.42
\pm 0.26$\,ps, which is equivalent to a biphoton coherence
bandwidth of $2.5 \pm 0.4$\,nm. The measured coherence bandwidth
is in excellent agreement with the 2.5\,nm bandwidth calculated
from PPKTP's Sellmeier equations \cite{epmSHG}. The HOM visibility
is defined by $V = 1-C_{\rm min}/C_{\rm max}$, where $C_{\rm min}$
is the coincident count at zero signal-idler time delay and
$C_{\rm max}$ is the coincident count at a long time delay (in the
flat part of the HOM measurement). The fiber-to-fiber coupling
efficiency of the air gap varied by $\sim$10\% over the scan, and
we normalize the coincidence data to the maximum singles count
rate in Figs.~\ref{cwhom} and \ref{pulsedhom}; this normalization
affects the visibility by $\sim$1\%.   The cw measurements
represent the first demonstration of tunable SPDC without any
degradation of the biphoton entanglement, a result due to the
large extended phase-matching bandwidth of 67\,nm.  The tuning
range of 10\,nm centered at the wavelength of 1584\,nm is 4 times
larger than the coherence bandwidth.  We measured that the FPBS
had an extinction ratio of $\sim$99\% for both ports. A 2\%
reduction in the visibility is caused by the 1\% leakage of the
signal field into the idler channel and vice versa. This 2\% loss
could be eliminated by using a clean-up polarizer in each of the
two output ports of the FPBS.

For the pulsed HOM measurement, the setup was identical to that
used in the cw measurements, except the pump laser was set in the
pulsed mode centered at 790\,nm with a 3-dB bandwidth of 6\,nm.
Figure~\ref{pulsedhom} shows the pulsed HOM measurement results
with a visibility of $85 \pm 7$\% and a biphoton coherence time
$\tau_c=1.3 \pm 0.3$\,ps.  This is the first observation of high
HOM visibility, without spectral filtering, in pulsed SPDC in
which the pump bandwidth $\Omega_p$ is much larger than the
biphoton coherence bandwidth $\Omega_c$. The visibility in pulsed
HOM is sensitive to deviations from the exact EPM operating
conditions such as the crystal angle (affecting the effective
grating period) and the pump wavelength. The slightly lower
visibility in pulsed HOM results compared with the cw measurements
may be caused by such deviations.  Another plausible explanation
for the lower visibility in pulsed HOM interferometry is the
presence of higher order terms in the phase matching function that
have not been considered in theoretical analyses
\cite{epmPRL,epmPRA}.

We also measured the autocorrelation of the signal and idler
light.  The PPKTP output light was sent through a polarizer to
pass only the horizontally polarized signal or the vertically
polarized idler photons, then rotated by 45$^\circ$ with a second
HWP before being coupled into the fiber interferometer.  The fiber
interferometer, as depicted in the setup of Fig.~\ref{setup},
became effectively a Mach-Zehnder interferometer because the FPBS
now served as a 50--50 beam splitter for the input signal or idler
light.  The autocorrelation measurement shows a signal coherence
time of 380\,fs, and an idler coherence time of 350\,fs which are
much shorter than the biphoton coherence time of $1.3 \pm
0.3$\,ps.

The high visibility obtained in our pulsed HOM measurements is the
consequence of operating under EPM which allows for the generation
of coincident frequency entanglement. The signal and idler outputs
are frequency entangled if, given the generic biphoton state $|
\psi \rangle_{bi} = \iint d\omega_s d\omega_i A(\omega_s,
\omega_i) | \omega_s \rangle| \omega_i \rangle$, the spectral
amplitude $A(\omega_s,\omega_i)$ cannot be factorized as
$f_s(\omega_s)f_i(\omega_i)$ \cite{wamsley2}.  If the two photons
are not entangled, then $S_s(\omega_s) = |f_s(\omega_s)|^2$ and
$S_i(\omega_i) = |f_i(\omega_i)|^2$ are the fluorescence spectra,
which are directly related to their coherence times.  If the two
interfering photons at the 50--50 coupler of the HOM setup were
not entangled, such as those from two single-photon sources, the
HOM coherence time would be given by the overlap integral of the
two pulses, as shown in Refs.\ \cite{HOM,Loudon,Yamamoto}. In our
case, the pulsed biphoton coherence time of 1.3\,ps is much longer
than the single photon coherence times which are less than
400\,fs. Such a long biphoton coherence time would not be possible
if the two photons were not entangled. The large difference
between the measured biphoton coherence time and the single photon
coherence times suggests a significant amount of frequency
entanglement. However, we believe that the output state was not
maximally entangled. Such a non-maximally entangled output state
is consistent with our observations: a small reduction in HOM
visibility and a slightly shorter HOM coherence time compared with
the cw results.  Further measurements and theoretical analyses are
needed to clarify the effects of various pump and crystal
parameters on the amount of entanglement and HOM visibilities.

In summary, we have demonstrated a new technique of generating
entangled photons with coincident frequencies in parametric
down-conversion under extended phase matching conditions.  HOM
interferometric measurements show high visibilities for cw or pulsed
pumping, valid for a broad tuning range or pump pulse bandwidth that
is less than the large extended phase matching bandwidth.  Pumping
the crystal bidirectionally (as shown in \cite{dual-pump}) under EPM
would allow us to generate photons that are both
coincident-frequency and polarization entangled. Pulsed entanglement
provided by this quantum state suits the needs of LOQC experiments
which require high timing resolution and high quality interference
\cite{KLM}. Also, timing and positioning measurements can be
improved by $\sqrt{2}$ over the standard quantum limit if
coincident-frequency entangled photons are used in time-of-arrival
measurements \cite{lloyd-nature}.

The authors would like to acknowledge J.\ Sickler and J.\ Gopinath
for help with the fiber interferometer and T.\ Kim for help with
coincidence detection circuitry. This work was supported by the
Advanced Research and Development Activity (ARDA), administered by
the Office of Naval Research under Grant N00014-03-1-0869.

\end{document}